\documentclass[11pt,twoside]{article}

\usepackage{asp2006}
\usepackage{epsf}
\usepackage{lscape}

\begin{document}
\title{Full Text Searching in the Astrophysics Data System}
\author{G\"unther~Eichhorn, Alberto~Accomazzi,
Carolyn~S.~Grant, Edwin~A.~Henneken, Donna~M.~Thompson,
Michael~J.~Kurtz, Stephen~S.~Murray}
\affil{Harvard-Smithsonian Center for Astrophysics, 60 Garden Street,
Cambridge, MA 02138}

\begin{abstract}
The Smithsonian/NASA Astrophysics Data System (ADS) provides a search
system for the astronomy and physics scholarly literature.  All major
and many smaller astronomy journals that were published on paper have
been scanned back to volume 1 and are available through the ADS free
of charge.  All scanned pages have been converted to text and can be
searched through the ADS Full Text Search System. In addition,
searches can be fanned out to several external search systems to
include the literature published in electronic form.  Results from the
different search systems are combined into one results list.

\medskip
\noindent
The ADS Full Text Search System is available at:

\medskip

http://adsabs.harvard.edu/fulltext\_service.html

\end{abstract}

\section{Introduction}

The Smithsonian/NASA Astrophysics Data System (ADS) provides access to
the astronomy and physics literature.  As of September 2006 it
provides a search system for almost 4.9 million records, covering most
of the astronomical literature (including planetary sciences and solar
physics) and a large part of the physics literature.  The ADS has been
described in detail in a series of articles in Astronomy and
Astrophysics Supplements
\citep{2000A&AS..143...41K,2000A&AS..143...61E,2000A&AS..143...85A,
2000A&AS..143..111G}.

Since 1994, the Astrophysics Data System (ADS) has scanned the
scholarly literature in astronomy.  As of September 2006, we have
scanned over 3.3 million pages.  These articles are available free of
charge world-wide.

In order to make this resource even more accessible, we have used
Optical Character Recognition (OCR) software to obtain the text of all
the scanned pages in ASCII form.  This allows us to index the full text
of all articles and make it searchable.

This search system covers the astronomical literature that was
published only on paper.  In order to search the literature published
in electronic form, we developed a system that sends queries to
the search systems of several publishers.  The results of these
queries are then combined with the results of the ADS internal queries
to seamlessly cover the majority of the literature.

This article describes some of the features of the search system for
the full text of the astronomical literature.

\section{Current Data in the ADS}

We have so far scanned about 3.3 million pages from 43 journals, 15
conference series and 67 individual conferences, as well as a
significant number of historical observatory publications.  The
scanned pages as of September 2006 use about 600 GB of disk space, the
OCRd text uses 72 GB.  The OCRd text is so-called ``dirty OCR''
because it has not been checked manually and it contains significant
numbers of errors.  This means that this text cannot for instance be
used to extract numerical data from tables, it would be inaccurate.
However, for searching the text for specific words, this ``dirty OCR''
is good enough.  Significant words are usually used more than once, so
even if the OCR software made a mistake in recognizing a word once, it
will still show up correctly in other places of the same article.

Indexing of the OCRd text proved to be challenging.  The number of
unique words from this text is large.  One reason for the large
number of words is the fact that mistakes during the OCR process
create new misspelled words.  To reduce this problem, we remove words
that have spurious characters in them that are OCR errors.
But even after removing such words, as well as other unusable words
like numbers, there are 14 million unique words in the index. The
files produced during indexing are large, the largest being about 3.7
GB, close to the limit of 32 bit addressing.  There is still some room
for growth, but eventually we will have to move to 64 bit addressing
for the full text search system in the ADS.

\section{Search Forms}

There are two search forms available.  The basic search form allows
you to enter the search term(s) and select which search systems to
query.  The search terms are combined with AND, meaning that all
search words must be present on a page in order to be selected.  The
system supports phrase searching when multiple words are enclosed in
double quotes.  By default, synonym replacement is enabled.  This
means that the system not only searches for a specified word
but also for all other words that have the same meaning.  Synonym
replacement can be turned off for individual words be pre-pending a
'='.  This will search for the exact spelling of the word, which can
be useful for words that have synonyms that are very common and would
produce many matches.  For instance ``galaxy'' is a synonym for
``extragalactic''. A search for ``=extragalactic'' will remove
``galaxy'' from the matches.

The advanced search form allows in addition the selection of a
publication date range and a journal.  It also allows the selection of
several sort options.  One important sort option is ``oldest first''.
This allows you to find the first occurrence of a word or phrase in the
literature (see \ref{ex} Example Usage).

\section{Returned Data}

The search returns a list of articles that contain the search terms.
Under each article it lists each page individually that contains the
search terms.  It includes a partial sentence around the search terms,
with the search term highlighted in red.  For pages that are not in
articles (cover pages, various other material, and pages from issues
where we don't have the pagination information), the pages are listed
individually.  The article information links back to the regular ADS
abstract service, the page information links directly to the scanned
page.

\section{Combined Searches}

In order to include the more recent literature that was published in
electronic form, the user can select to include one or more external
search systems in the query.  The external search systems are queried
in parallel.  As results are returned from the external systems, they
are displayed to the user.  Once all results are available, a final
combination of all the results is compiled and displayed.  The search
fan-out is still experimental.  It is not yet very stable since none
of the external systems provide a dedicated interface for such
external queries.  It was implemented by simulating regular user
queries to these systems.  This makes our fan-out system vulnerable to
changes in the external search systems.  If an API (Application
Programming Interface) becomes available for any of the external
systems, we will implement it to build a more stable system.

We currently query the systems listed in table~\ref{ss}.

\begin{table}[!ht]
\caption{External Search Systems}
\label{ss}
\smallskip
\begin{center}
{\small
\begin{tabular}{ll}
\tableline
\noalign{\smallskip}
Search System & Journals searched\\
\noalign{\smallskip}
\tableline
\noalign{\smallskip}
Google Scholar & Monthly Notices of the Royal Astronomical Society\\
 & Annual Review in Astronomy and Astrophysics\\
 & Annual Review of Earth and Planetary Sciences\\
 & Applied Optics\\
 & Journal of the Optical Society of America\\
University of Chicago Press & Astronomical Journal\\
 & Astrophysical Journal\\
 & Astrophysical Journal Letters\\
 & Astrophysical Journal Supplement\\
 & Publications of the Astronomical Society\\
 & of the Pacific\\
EDP Sciences & Astronomy and Astrophysics\\
Nature & Nature\\
 & Nature Physics\\
 & Nature Physical Science\\
National Academy of Science & Proceedings of the National Academy of Science\\
\noalign{\smallskip}
\tableline
\end{tabular}
}
\end{center}
\end{table}

\section{\label{ex} Example Usage}

Using the full text search system is different from using the abstract
search system in the ADS.  Since there are so many more words in the
full text, there are usually many more matches.  It is therefore
generally advisable to use more unique words, more search terms, and/or
phrases.

For instance if you are trying to find out when the concept of a
critical mass was first described, searching for the words ``critical
mass'' without the double quotes would not produce anything useful,
but a search for the phrase

\smallskip
``critical mass''

\smallskip
\noindent
with double quotes from the Advanced Search form, with ``Oldest
first'' selected, quickly finds an article in PASP from 1919 that
attributes this phrase to Professor Eddington.

Another interesting question is to find out when the name Pluto was
first suggested for a planet.  If you enter:

\smallskip
planet pluto

\smallskip
\noindent
in the search field and select ``Oldest first'' under the sort
options.  One of the first matches will be of an article in ``The
Observatory'' from 1898 that suggests using Pluto as the name for the
recently discovered planet DQ.  Incidentally, the name had to wait for
another 30 years before it was actually used for a planet.  This
capability can be very useful for astronomy historians.

\section{Conclusion}

The ADS provides a search capability for the full text of a large part
of the astronomical literature.  This capability complements the
regular abstract search system.  It allows the in-depth analysis of
the older literature and especially the historical observatory
publications, a part of the astronomical literature that has not
been accessible in any search system until now.

\acknowledgements

The ADS is funded by NASA Grant NNG06GG68G.

\end{document}